\documentclass[%
 reprint,
 amsmath,amssymb,
 aps,
 showkeys
]{revtex4-2}

\usepackage{graphicx}% Include figure files
\usepackage{dcolumn}% Align table columns on decimal point
\usepackage{bm}% bold math
\usepackage[T1]{fontenc}
\usepackage{xcolor}
\usepackage{multirow}
\usepackage{url}

\DeclareUnicodeCharacter{0229}{\l{}}
\DeclareUnicodeCharacter{202F}{\textendash}

\begin{document}

\title{Bandgap manipulation of hBN by alloying with aluminum: absorption properties of hexagonal BAlN}

\author{Jakub Iwański}
\email{Jakub.Iwanski@fuw.edu.pl}
\author{Mateusz Tokarczyk}
\author{Aleksandra K. Dąbrowska}
\author{\\Jan Pawłowski}
\author{Piotr Tatarczak}
\author{Johannes Binder}
\author{Andrzej Wysmołek}
\affiliation{Faculty of Physics, University of Warsaw, ul. Pasteura 5, 02-093 Warsaw, Poland}
%\date{\today}

\begin{abstract}

The versatile range of applications for two-dimensional (2D) materials has encouraged scientists to further engineer the properties of these materials. This is often accomplished by stacking layered materials into more complex van der Waals heterostructures. A much less popular but technologically promising approach is the alloying of 2D materials with different element compositions. In this work, we demonstrate a first step in manipulating the hBN bandgap in terms of its width and indirect/direct character of the optical transitions. We present a set of aluminum alloyed hexagonal boron nitride (hBAlN) samples that were grown by metal organic vapor phase epitaxy (MOVPE) on 2-inch sapphire substrates with different aluminum concentration. Importantly, the obtained samples revealed a sp$^2$-bonded crystal structure. Optical absorption experiments disclosed two strong peaks in the excitonic spectral range with absorption coefficient $\alpha \sim 10^6$~cm$^{-1}$. Their energies correspond very well with the energies of indirect and direct bandgap transitions in hBN. However, they are slightly redshifted. This observation is in agreement with predictions that alloying with Al leads to a decrease of the bandgap energy. The observation of two absorption peaks can be explained in terms of mixing electronic states in the K and M conduction band valleys, which leads to a significant enhancement of the absorption coefficient for indirect transitions.

\end{abstract}


\maketitle

\section{INTRODUCTION}

The initial research on graphene has motivated researchers to explore the properties of other atomically thin layered materials (2D materials) and their stacks, sometimes referred to as NanoLego structures \cite{novoselov2004, novoselov2016, gupta2015, gibertini2019}. This is typically accomplished by stacking different layers together into into van der Waals heterostructures \cite{guo2021, liang2020}. Combining these materials using epitaxial growth provides the additional opportunity to scale-up such structures or to create new ones by growing materials as alloys with appropriate composition \cite{yao2022, singh2021}. In the case of hexagonal boron nitride (hBN), the natural choice is alloying hBN with aluminum and gallium, since they are in the same group of the periodic table as boron. Research on the novel hB$_{1-x-y}$Al$_x$Ga$_y$N material (where $x$ and $y$ stand for Al and Ga concentration in the alloy respectively) is interesting in terms of exploration of new physical phenomena as well as form a technological point of view. hBN with its, unusual for III-N, strong sp$^2$ covalent bonds in plane and weak van der Waals bonds out of plane exhibits extraordinary properties \cite{zunger1976}. One of the intriguing facts is that inspite of the indirect character of the bandgap, hBN shows a photoluminescence (PL) emission intensity of the band-edge transition as strong as for direct semiconductors \cite{elias2021}. Thus, the possibility of tuning the hBN bandgap using Al opens up further prospects for applications in hBN-based quantum wells and optoelectronic devices. The same scheme could be used for hBN alloying with Ga. Tunable hBAlN, hBGaN or hBAlGaN alloys could become promising candidates for building blocks of efficient light sources in the deep UV spectral range (DUV, 200-280~nm, 4.4-6.2~eV). Such candidates are of high importance, since nowadays the efficiency of semiconductor light sources operating in DUV is very limited ($\sim$10\% for 270-280~nm and $\sim$3\% for 250~nm) \cite{amano2020}. This spectral range is crucial, for example for the disinfection and sterilization of air, water and surfaces \cite{ramos2020, chen2017}.

However, before hBN-based quantum structures and DUV light sources can be realized, more fundamental questions must be answered. Firstly, it is still unclear how to tune the hBN bandgap energy. This knowledge would be of great importance for designing hBN-based quantum wells that would trap carriers and enhance emission from the structure. Moreover, it would be very beneficial for further improvement if one could change the nature of the hBN bandgap from indirect to direct \cite{Zhang2017, turiansky2019}.

In this work, we present results for boron nitride layers grown by metal organic vapor phase epitaxy (MOVPE) that were alloyed with aluminum. We obtained materials that preserve a sp$^2$-bonded crystal structure. The produced hB$_{1-x}$Al$_x$N layers with aluminum concentration $x$ up to a few percent exhibit strong absorption for two energies that coincide with energies of indirect and direct bandgap in hBN. Our approach that uses MOVPE to grow this novel material provides a perspective for further development of the hBAlN alloy composition control in terms of doping for conductivity, manipulation of the bandgap, and commercial up-scaling.

\section{METHODS}
\subsection{Samples}

The samples used in these studies were grown using an Aixtron CCS 3$\times$2'' metal organic vapor phase epitaxy system. The growth was carried out on 2-inch sapphire $c$-plane substrates with 0.3$^\circ$ off-cut. For the growth of layered sp$^2$-bonded hB$_{1-x}$Al$_x$N samples ammonia, triethylboron (TEB) and trimethylaluminum (TMAl) were used as nitrogen, boron, and aluminum precursors, respectively and hydrogen was used as a carrier gas. Note that we used the hBAlN notation (with letter h) to highlight the importance of the hexagonal hBN-like crystal structure of the obtained material. The precursors were injected alternatively following the flow modulation epitaxy growth mode \cite{kobayashi2008}. The scheme of pulses in a single cycle was as follows: TEB-NH$_3$-TMAl-NH$_3$. The growth temperature was kept at 1300~$^\circ$C. The temperature value was obtained with an in situ multipoint optical pyrometer (ARGUS). NH$_3$ and TEB flows were fixed for all samples (V/III ratio for TEB-NH$_3$ pulses was $\sim 15 000$). The TMAl/III ratio (III = TEB + TMAl) in the process varied for the samples: $S_{0.02}$ (ratio 0.02), $S_{0.04}$ (ratio 0.04), $S_{0.07}$ (ratio 0.07), $S_{0.13}$ (ratio 0.13). Such a ratio (TMAl/III) gives an insight into relation between B and Al source in a gas phase during the process and very initial predictions of the material composition $x$. To obtain the lowest ratio, the duration of consecutive pulses of TMAl and NH$_3$ was reduced by half. This step was enforced by the minimal precursor flow limitation in our system. The number of pulsing cycles was chosen in such a way that all samples were grown 60 minutes.

\subsection{Experimental details}

The crystal structure of the samples was examined with a Panalytical X'Pert diffractometer with standard CuK$_\alpha$ x-ray source. The x-ray light beam was formed by an x-ray mirror for Bragg-Brentano optics. Infrared reflectance spectra were collected with a Thermo Scientific Nicolet Continuum Infrared Microscope equipped with a 32x Schwarzschild infinity corrected objective (NA 0.65). All samples of $7\times10$~ mm were measured with a perpendicular incident beam on five $70\times70$~$\mu$m areas placed in the center and four corners of the sample. Spectra were collected in the range of 650-4000~cm$^{-1}$ with a resolution of 0.482~cm$^{-1}$. The surface morphology was examined by scanning electron microscopy (SEM) using a FEI Helios NanoLab 600 system. Absorption spectra were measured using a Varian Cary 5000 UV-vis-NIR spectrophotometer in dual-beam mode with nitrogen purging. The spectral bandwidth was set to 1~nm.

Although our samples are very homogeneous on the wafer scale \cite{Dabrowska2020, ludwiczak2021}, we decided to measure each of the samples taken from the exact same area on the wafer.

\section{RESULTS}

\subsection{x-ray diffraction}

\begin{figure}[b]
\includegraphics[width=\columnwidth]{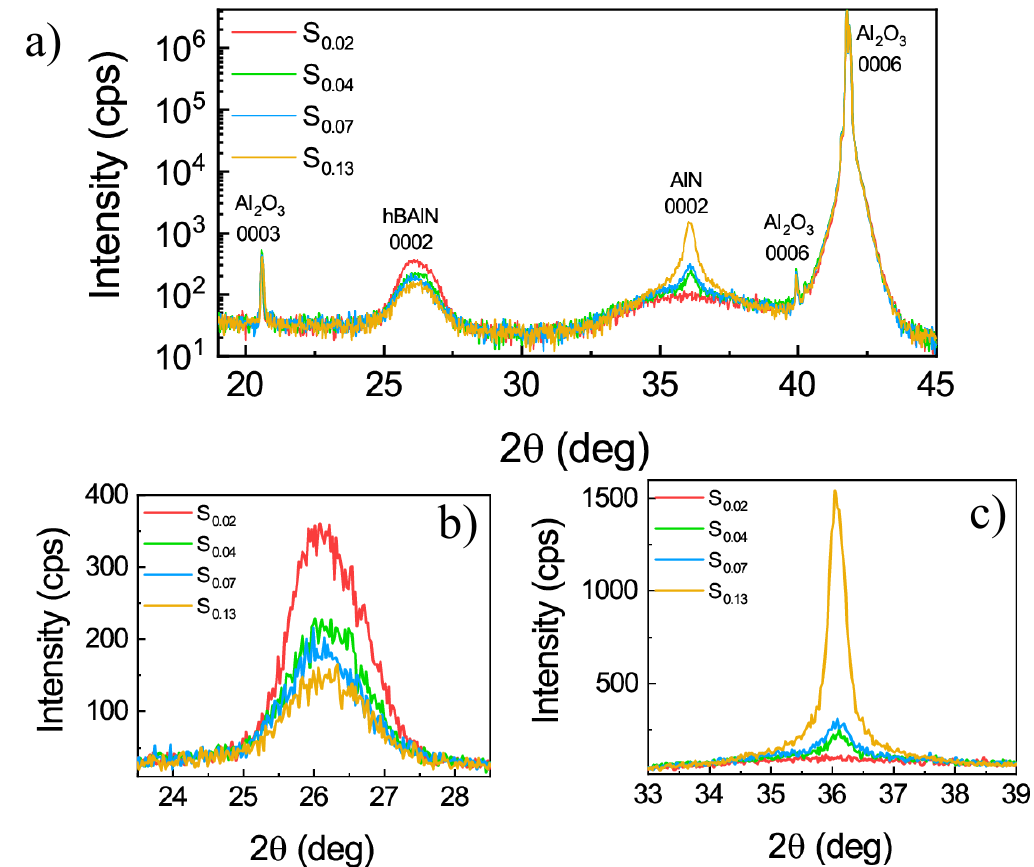}
\caption{\label{fig:XRD} X-ray diffraction $2\theta/\omega$ scans of the studied samples: a) scan in a broad $2\theta$ range, b) zoom on the peak related to hBAlN and c) zoom on the peak related to AlN.}
\end{figure}

In figure~\ref{fig:XRD}a) we present x-ray diffraction $2\theta/\omega$ scans collected for all the studied samples. The peaks at $\sim 20.5^\circ$, $\sim 40^\circ$ and $\sim 41.75^\circ$ correspond to the 0003 and 0006 planes of Al$_2$O$_3$. In figures~\ref{fig:XRD}b),~c) we present a zoom on the peaks at $\sim 26^\circ$ and $\sim 36^\circ$ that come from the reflection of the 0002 planes in sp$^2$-BAlN and sp$^3$-AlN. The parameters of the Gaussian curve fit performed on the data concerning hBAlN are included in table~\ref{tab:XRD}. Note that the 0002 BAlN peak related to sample $S_{0.02}$ is asymmetric, so we fitted two Gaussian curves yielding two different components of the peak. The components originate from turbostratic (tBAlN, lower angles) and hexagonal (hBAlN, higher angles) phases of sp$^2$-bonded BAlN \cite{Dabrowska2020, Tokarczyk2023}. In the further analysis we will focus only on the hexagonal phase. In table~\ref{tab:XRD} we do not observe significant changes in the peak position $2\theta_B$ and the full width at half maximum (FWHM) for the hBAlN peak. This implies that the $c$ lattice constant is comparable for all the samples ($\sim 3.41$~\AA) as well as the thickness of the crystal. The only monotonous trend can be found in the amplitude of the hBAlN peak. This means that the misorientation of the 0002 crystal planes is enlarged when increasing the TMAl flow.

\begin{table}[htbp]
\caption{\label{tab:XRD}
Parameters of the Gaussian curve fit to 0002 BAlN XRD peak.}
\begin{ruledtabular}
\begin{tabular}{lccc}
Sample & amplitude (cps) & 2$\theta_B$ (deg) & FWHM (deg) \\
\hline
\multirow{2}{*}{$S_{0.02}$} & 7.0(6) & 25.84(1) & 0.51(4)  \\
            & 28.2(5) & 26.187(9) & 1.28(1) \\
$S_{0.04}$ & 18.5(2) & 26.136(6) & 1.29(1)  \\
$S_{0.07}$ & 15.0(2) & 26.078(7) & 1.31(2)  \\
$S_{0.13}$ & 11.4(2) & 26.122(8) & 1.36(2)  \\
\end{tabular}
\end{ruledtabular}
\end{table}

The signal associated to AlN consists of a very broad peak, which is alike for all the samples and a narrower component that changes its intensity. In fact, these two components have different origins. The broad component comes from an unintentional thin ($\sim$2~nm) AlN layer that is created on the hBAlN-sapphire interface during the initial growth stage. This kind of layer is characteristic for MOVPE-grown hBN \cite{Tokarczyk2023, Iwanski2021}. However, in accordance with the enhancement of the sharp component with an increase in TMAl, we can deduce the formation of AlN crystals in our samples. These crystals do not seem to influence the crystal structure of hBAlN as we do not observe any correlation between the intensity of the sharp peak at $\sim 36^\circ$ and the values of $2\theta_B$ and FWHM in table~\ref{tab:XRD} which are related to BAlN.

\subsection{Fourier-transform infrared spectroscopy}

\begin{figure}[b]
\includegraphics[width=\columnwidth]{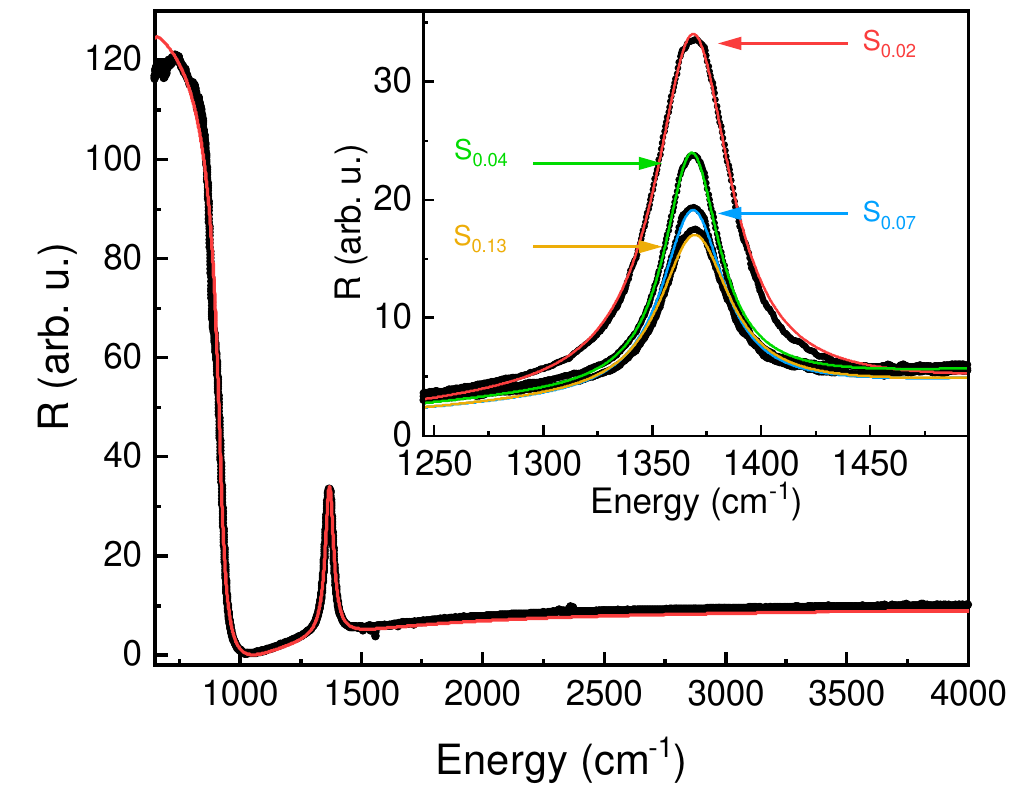}
\caption{\label{fig:FTIR} Fourier-transform infrared reflectance spectrum for the $S_{0.02}$ sample with the lowest TMAl flow (black dots). The red line is the fitted curve. The inset shows a zoom on the peak of the $E_{1u}$ vibrational mode, which is strong evidence for a sp$^2$ crystal structure for all hBAlN samples. The black dots represent the experimental data and the lines are the fitted curves: red for $S_{0.02}$, green for $S_{0.04}$, blue for $S_{0.07}$, yellow for $S_{0.13}$.}
\end{figure}

In figure~\ref{fig:FTIR} we present exemplary infrared reflectance spectra obtained for the studied samples. The peak observed about 1368~cm$^{-1}$ is related to the $E_{1u}$ vibrational mode characteristic for hBN \cite{geick1966}. The observation of this peak provides direct evidence that our samples are sp$^{2}$ bonded. The high reflectance below 1000~cm$^{-1}$ comes from the sapphire substrate. Barely visible small and sharp peaks about 1550~cm$^{-1}$, 2300~cm$^{-1}$ and 3500~cm$^{-1}$ correspond to transitions in the atmospheric gases present during the experiment.

To make the FTIR results quantitative, we implemented a spectra analysis method, as described in the Ref.~\cite{iwanski2022}. In this method, it is assumed that the material is composed of harmonic oscillators with self-energy $\omega$ and damping parameter $\gamma$. It makes use of the Dynamic Dielectric Function (DDF) of the materials in the structure (boron nitride and sapphire in this case). The substantial advantage of this method is the possibility of characterizing the grown layer itself. The self-energy of the oscillator $\omega_{BAlN}$ (phonon energy, peak position), its damping parameter $\gamma_{BAlN}$ (peak broadening) and layer thickness provide information about strain, homogeneity and growth rate. In table~\ref{tab:FTIR} we present the best fit parameters obtained as an average of 5 points measured for each sample.

\begin{table}[htbp]
\caption{\label{tab:FTIR}
Averages of the best fit parameters of the FTIR spectra obtained for the studied samples.}
\begin{ruledtabular}
\begin{tabular}{lccc}
Sample&$\omega_{BAlN}$ (cm$^{-1}$)&$\gamma_{BAlN}$ (cm$^{-1}$)&$d_{BAlN}$ (nm)\\
\hline
$S_{0.02}$ & 1368.94(5) & 24.5(1) & 18.6(5)  \\
$S_{0.04}$ & 1368.03(3) & 22.1(2) & 9.1(4)  \\
$S_{0.07}$ & 1368.7(2) & 26.5(6) & 9.8(3)  \\
$S_{0.13}$ & 1369.3(2) & 30.1(5) & 10.1(3)  \\
\end{tabular}
\end{ruledtabular}
\end{table}

The fitted parameters are monotonous for the samples $S_{0.04}$, $S_{0.07}$ and $S_{0.13}$. The increase in TMAl flow leads to an increase in phonon energy and peak broadening, which suggest the introduction of an inhomogeneous compressive strain within the layer. The changes of $\omega_{BAlN}$ ($\sim$0.63~cm$^{-1}$ for the following samples) and $\gamma_{BAlN}$ ($\sim$4~cm$^{-1}$ for the following samples) are significant in terms of FTIR measurements. Remarkably different is the trend for the layer thickness. The 4-time increase in amount of TMAl (samples $S_{0.04}$ and $S_{0.13}$) resulted in a layer that is only 1~nm thicker, which is equivalent to the increase of the growth rate by less than 0.02 nm/min.

The sample $S_{0.02}$ clearly stands out and does not follow the same trends. The reason for this behavior is most likely due to the change of TMAl-NH$_3$ pulse duration necessary to obtain a TMAl/III ratio equal to 0.02. This led to a double increase in growth rate compared to the sample $S_{0.04}$. This result is reasonable, since the TMAl/III ratio for $S_{0.02}$ was twice higher than for $S_{0.04}$. At the same time the number of TEB pulses increased by 33\%. So we find the competition between TEB-NH$_3$ and TMAl-NH$_3$ pulses to be more important for the growth rate rather than  the change in number of TEB pulses in the growth process.

\subsection{scanning electron microscopy}

\begin{figure}[t]
\includegraphics[width=\columnwidth]{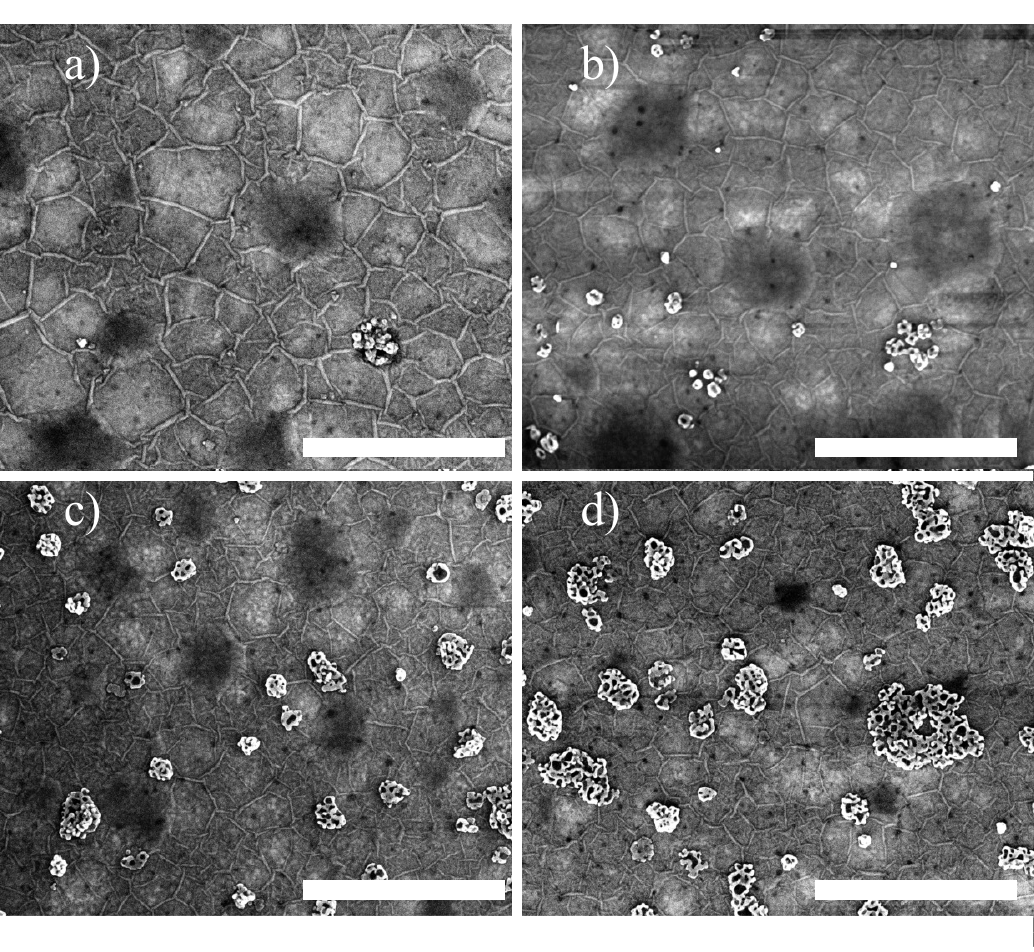}
\caption{\label{fig:SEM} Scanning electron microscopy images of the samples a) $S_{0.02}$, b) $S_{0.04}$, c) $S_{0.07}$, d) $S_{0.13}$. The scale bar corresponds to 1~$\mu$m for each image.}
\end{figure}

The morphology of the studied samples is presented in figure~\ref{fig:SEM}. For each sample, a characteristic wrinkle pattern is observed. The wrinkles are created during cooling down the material after the growth process and are due to the relaxation of strain caused by the difference in thermal expansion coefficients of the layer and the substrate \cite{iwanski2022}. Because of the mechanism of the wrinkle formation their presence is an evidence for the continuity of the hBAlN epitaxial layer. The wrinkles are most pronounced for sample $S_{0.02}$, which is due to a much larger thickness. The darker circular areas in the images are bubbles with hydrogen inside. They are created during SEM imaging and are the result of the radiolysis of interfacial water via electron irradiation \cite{binder2023}. The size of crystalline objects on the surface of the layer is correlated with the amount of TMAl present in the process. This observation is in good agreement with the XRD results for which the AlN narrow diffraction peak was observed for $S_{0.04}$, $S_{0.07}$ and $S_{0.13}$. This leads to the conclusion that the observed crystals are related to sp$^3$ bonded aluminum nitride. More detailed atomic force microscopy and Raman spectroscopy studies of these objects are presented in the Supplementary Materials.

\subsection{UV-vis spectroscopy}

\begin{figure}[t]
\includegraphics[width=\columnwidth]{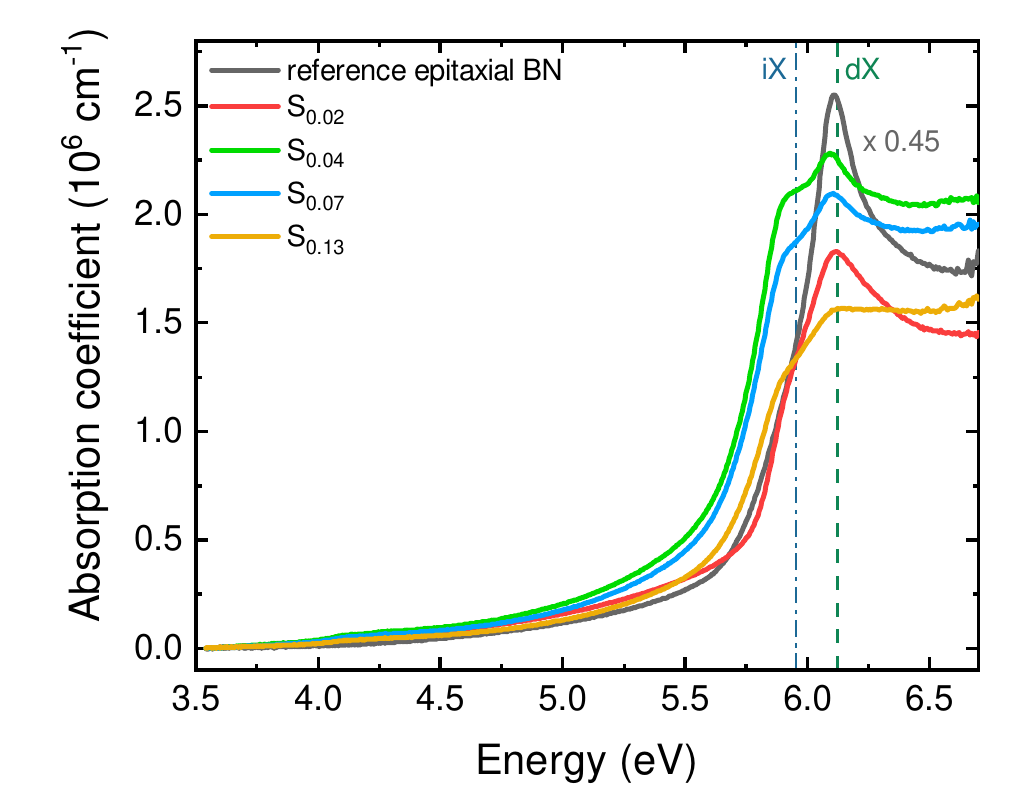}
\caption{\label{fig:Abs} Absorption coefficient spectra of the samples studied. The solid gray line is the spectrum collected for the reference boron nitride sample without aluminum. It was multiplied by factor 0.45 for clarity. The reference sample was grown in analogy to the $S^{I}$ from Ref.~\cite{Dabrowska2020}. The uncertainty of the results is of order of $0.5\times 10^5$~cm$^{-1}$. Green dashed and dash-dotted navy lines illustrate the energies of direct and indirect bandgaps in boron nitride.}
\end{figure}

To extract the absorption coefficient from the measured absorbance, we subtracted the absorbance spectrum obtained for bare sapphire. To calculate the absorption coefficient we used the thickness of the layer $d_{BAlN}$ taken from table~\ref{tab:FTIR}. The absorption coefficient spectra of the samples studied are presented in figure~\ref{fig:Abs}. The spectrum for the reference epitaxial boron nitride without aluminum sample was obtained in the same way. The reference sample was grown analogously to the sample $S^I$ presented in the work in Ref.~\cite{Dabrowska2020}. The absorption coefficient for all the samples is of order of $\alpha=2\times 10^6$~cm$^{-1}$. This value is twice higher as compared to those previously reported for boron nitride \cite{zunger1976, li2016, li2012}. In contrast to the reference sample, the samples with aluminum exhibit two well-resolved peaks. In previous studies only peak shifting was observed, which was accompanied by a broadening due to a decrease of sample quality \cite{Vuong2017}. In the case of our samples the two peak energies are close to the dash-dotted navy and dashed green lines that are positioned at energies corresponding to the indirect (5.955~eV \cite{Cassabois2016}) and direct (6.125~eV \cite{Watanabe2009, Rousseau2021}) band-edge transitions in boron nitride, respectively. Depending on the TMAl flow, one can change the intensity of those peaks and the intensity ratio between higher and lower energetic peaks. However, this relationship does not correlate with the amount of TMAl in the growth process.

The lower energy peak, which is not observed for the reference sample, could be thought to be the result of the crystaline objects presented in the SEM images (figure~\ref{fig:SEM}). However, the peak intensity does not scale with the number of objects on the layer's surface. Secondly, as has been mentioned before, the crystalities are identified as sp$^3$ bonded AlN which is known to have a bandgap energy of 6.2~eV \cite{perry1978}. Possible alloying of sp$^3$ bonded wurtzite-AlN (wAlN) with boron would, on the other hand, further increase the bandgap since wurtzite-BN (wBN) is a wider bandgap semiconductor \cite{Zhang2017}. Consequently, the signal coming from AlN should be observed in the absorption spectra for higher, not lower energies when comparing to the peak of pure boron nitride. However, we cannot provide conclusive information about the AlN-related spectral component since this energy range is close to the limit of detection for our spectrometer.

\section{DISCUSSION}

As presented in figure~\ref{fig:XRD}b) and table~\ref{tab:XRD}, the x-ray diffraction peak is observed at angles lower than expected for the 0002 hBN plane (26.764$^\circ$ \cite{joint_hBN}). This indicates a larger lattice constant in the $c$ direction. However, the peak position does not change significantly between the samples, which is in contrast to the results reported in Ref.~\cite{Vuong2020} in which authors observed peak position shifting towards higher angles with an increasing TMAl flow. According to theoretical DFT calculations presented in Ref.~\cite{GhorbanzadehAhangari2017}, a decrease in the $c$ lattice constant should be observed when hBN is alloyed with Al, since hexagonal aluminum nitride has calculated a smaller $c$ lattice constant. We conclude that since we do not have a perfect hexagonal phase the main reason for the XRD peak position change in our samples is that it is attributed to random twists of subsequent atomic layers, which make the material more turbostratic. In this case we postulate that the peak position variations related to the alloying with small amounts of Al are just a higher order correction. This hypothesis seems to be confirmed by the fact that all samples were grown under the same growth conditions. Another prominent feature presented in figure~\ref{fig:XRD}c) is the peak at $\sim 36^\circ$, whose intensity increases with the increase in TMAl flow. The value of the peak position is in agreement with the XRD peak of 0002 AlN in the crystal structure of wurtzite \cite{joint_AlN}. This indicates the creation of sp$^3$ bonded wAlN crystals, which scale with the amount of TMAl. This additional notable peak is a proof of the phase separation of sp$^2$ bonded hBAlN and sp$^3$ bonded wAlN. The observation agrees with SEM (figure~\ref{fig:SEM}), Raman spectroscopy and AFM (Supplementary Materials) measurements. Indeed, the number and size of crystaline objects correlate with the amount of TMAl, which provides strong evidence for the very limited solubility of aluminum in the hexagonal boron nitride layer at the temperature and pressure used in the growth process (1300~$^\circ$C, 400~mbar). Consequently, the experimental determination of the material composition is very difficult and needs further studies. Standard composition-determination techniques such as energy-dispersive x-ray spectroscopy (EDS), x-ray photoelectron spectroscopy (XPS), secondary ion mass spectroscopy (SIMS) cannot be employed to correctly determine the composition. Because the analyzed material is dielectric, electron charging is significant and does not allow EDS to be measured in one area for a longer period of time. Additionally, due to the electrons penetration depth in EDS it is hard to differentiate signal for Al that comes from the hBAlN layer and Al$_2$O$_3$ substrate. XPS and SIMS hardly distinguish Al from the hBAlN layer and AlN crystalities on its surface.

Although AlN crystalines can be observed, the sp$^2$-bonded, layered structure of hBAlN is maintained, which is proved by FTIR measurement presented in figure~\ref{fig:FTIR}. However, by introducing TMAl we modify the optical properties of the layer. This can be seen as a shift of the phonon energy $\omega_{BAlN}$ towards higher energies that suggests compressive stress in the structure as demonstrated in Ref.~\cite{androulidakis2018}. Furthermore, the increase in TMAl is followed by a broadening of the peak described by the parameter $\gamma_{BAlN}$, which indicates a defect-related inhomogeneity of the strain within the layer.

The most striking result that emerges from our work is the observation of an additional low energy peak in the absorption spectra. The hBN conduction band is known to consist of minima at the K and M points of the Brillouin zone that are energetically close to each other \cite{elias2021}. They are responsible for direct and indirect band transitions, respectively. As an indirect transition is a three-particle (photon, electron, phonon) event, it has a much lower probability to occur than a direct transition, which requires only two particles (photon, electron). Consequently, the absorption coefficient related to indirect transitions is usually 2-3 orders of magnitude lower as compared to direct ones. However, in the case of our samples, we observe two peaks at energies which coincide with the energies of direct and indirect transitions in hBN (dash-dotted and dashed lines in figure~\ref{fig:Abs}). Furthermore, both have a very high value of the absorption coefficient ($\alpha \sim 10^6$~cm$^{-1}$) typical for direct bandgap transitions. Notably that the lower-energy peak has even lower energy than expected for the indirect transition in hBN. This observation is in agreement with predictions about a decrease in bandgap energy when hBN is alloyed with Al \cite{Zhang2022}. To understand the mechanism that stands behind the observation of the two peaks in the absorption spectra, we need to be aware of the role of aluminum in the crystal structure. Al incorporation introduces short range disorder, which in turn would lead to the mixing of states with large $k$-vectors. As presented in Ref.~\cite{elias2021}, conduction and valence bands along KH and ML points in hBN are very flat and almost parallel to each other. This feature leads to high oscillator strength and consequently very efficient absorption. A disorder induced by Al incorporation further modifies the band structures and the probability of electronic transitions, i.e. oscillator strength. Consequently, other absorption channels are enabled through the defect-related states. The modification of states caused by Al-related defects allows us to observe highly effective absorption for both energies. However, further increase of TMAl flow and limited solubility of Al in hBN lead to a deterioration of optical quality of the material which is observed in broadening of spectroscopic peaks in the spectra (same for FTIR and UV-Vis).

\section{CONCLUSIONS}

In this work, we have shown the results for MOVPE grown hB$_{1-x}$Al$_x$N layers with TMAl/III ratio ranging from 0.02 to 0.13. X-ray diffraction and Fourier-transform infrared spectroscopy measurements proved that the obtained material maintained a sp$^2$-bonded layered crystal structure, which is characteristic for hBN. At this stage of the research, we are faced with a challenge of insufficient information regarding the exact composition of the material, as well as the presence of crystaline sp$^3$-bonded AlN clusters on the surface of the material. However, despite these obstacles, we managed to observe a significant change in the layer properties. Most importantly, we have shown two peaks of strong absorption ($\alpha \sim 10^6$~cm$^{-1}$) typical for direct excitonic transitions in hBN. The peak energies coincide with the energy of the indirect and direct bandgap transition in hBN. This observation becomes possible due to an activation of the absorption channels through defects caused by Al incorporation. The presented results are of great importance to understand how the boron nitride bandgap can effectively be manipulated in terms of its indirect/direct nature and its width. Understanding these processes is a key aspect for the fabrication of hBN-based structures for efficient deep UV emission.

\begin{acknowledgments}
This work was supported by the National Science Centre, Poland, under the decisions 2019/33/B/ST5/02766, 2020/39/D/ST7/02811 and 2022/45/N/ST5/03396.
\end{acknowledgments}

\bibliographystyle{vancouver}
\bibliography{bibliography}%

\end{document}

\beginsupplement

\newpage
\section*{AFM characterization of the crystaline objects}

Atomic force microscopy (AFM) experiments were performed using Dimension Icon microscope integrated with a NanoScope 6 controller (Bruker Corporation, Billerica, MA, USA). The topography images were collected in Peak Force Tapping\textsuperscript{\texttrademark} mode using Bruker RTESPA-300 probes made of antimony n-doped silicone (nominal spring constant of 40 N/m and a resonance frequency of 300 kHz). Nanomechanical mapping was performed using Peak Force Quantitative NanoMechanics\textsuperscript{\texttrademark} (QNM) mode. Before each experiment the exact value of spring constant was determined by Sader method. Moreover, deflection sensitivity of the cantilevers and curvature radius of the probes were evaluated using dedicated calibration samples and procedures. All measurements were recorded in air at constant temperature of $22 \pm 1$~$^\circ$C.

\begin{figure}[hbtp]
\includegraphics[width=\textwidth]{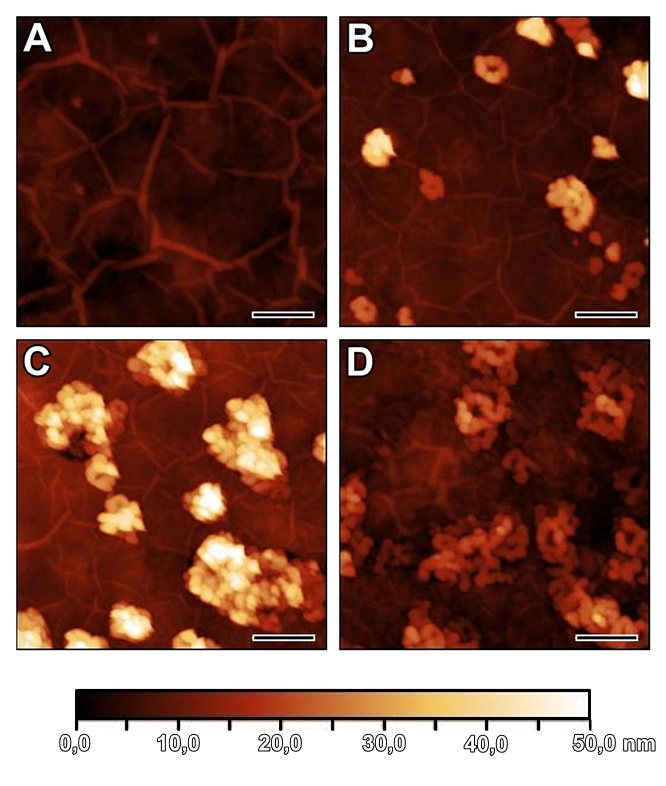}
\caption{\label{fig:AFM_topo}AFM topography images of hBAlN layers grown on sapphire substrate. a) $S_{0.02}$; b) $S_{0.04}$; c) $S_{0.07}$; d) $S_{0.13}$. The scale bars correspond to 200 nm.}
\end{figure}

\begin{figure}[hbtp]
\includegraphics[width=\textwidth]{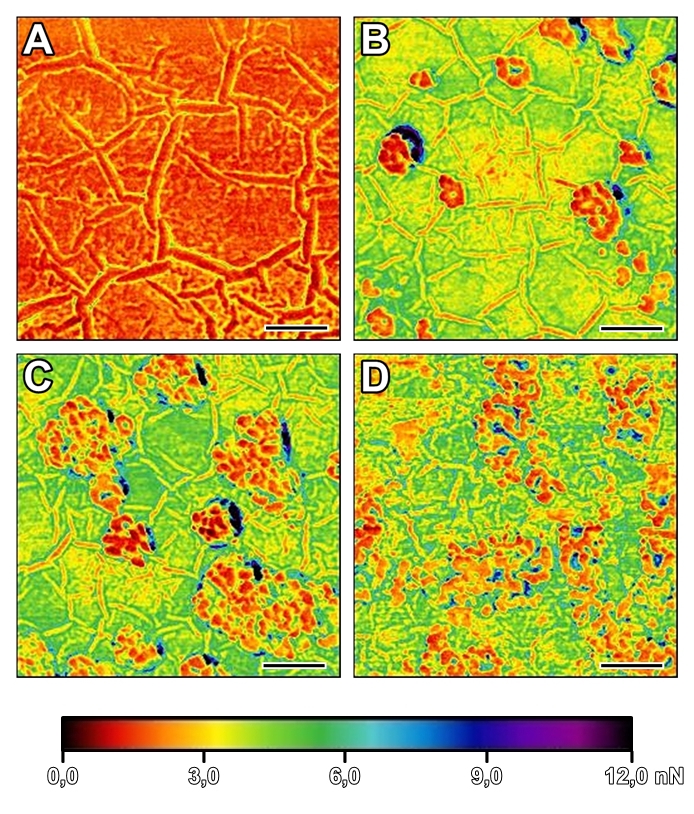}
\caption{\label{fig:AFM_adh}AFM adhesion images of hBAlN layers grown on sapphire substrat. a) $S_{0.02}$; b) $S_{0.04}$; c) $S_{0.07}$; d) $S_{0.13}$. The scale bars correspond to 200 nm.}
\end{figure}

\begin{figure}[hbtp]
\includegraphics[width=\textwidth]{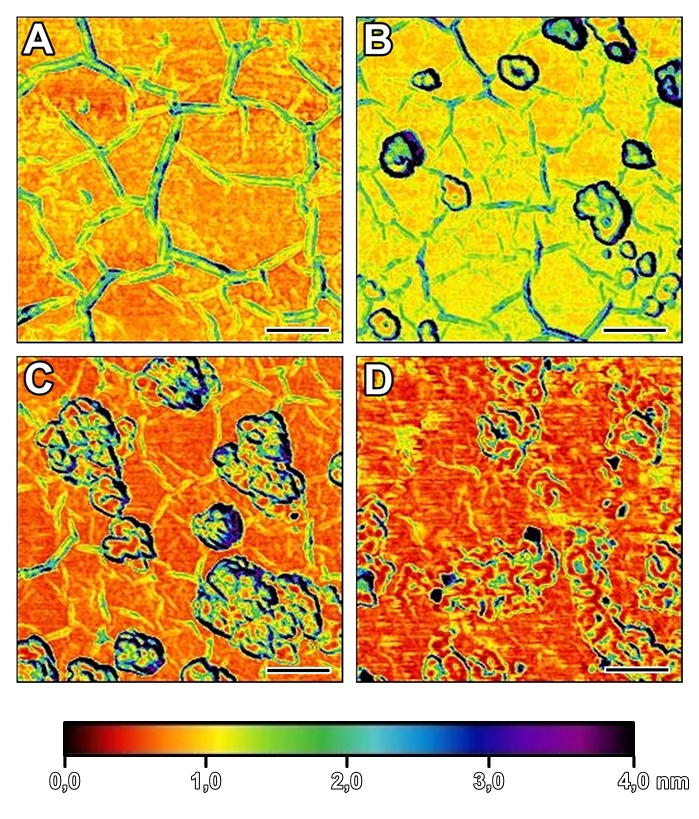}
\caption{\label{fig:AFM_defor}AFM deformation images of hBAlN layers grown on sapphire substrat. a) $S_{0.02}$; b) $S_{0.04}$; c) $S_{0.07}$; d) $S_{0.13}$. The scale bars correspond to 200 nm.}
\end{figure}

\begin{figure}[hbtp]
\includegraphics[width=\textwidth]{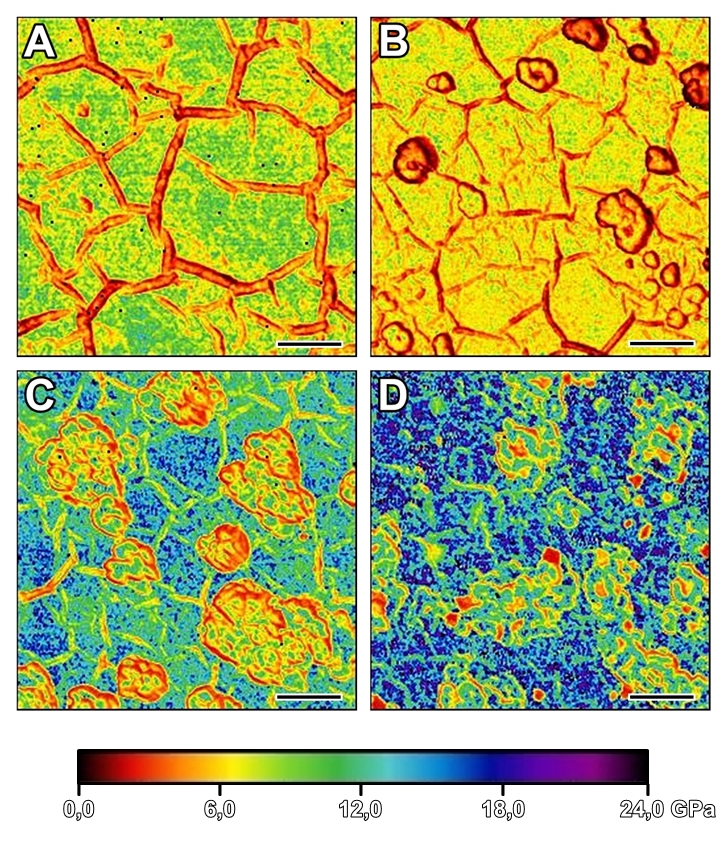}
\caption{\label{fig:AFM_stiff}AFM Young's modulus images of hBAlN layers grown on sapphire substrat. a) $S_{0.02}$; b) $S_{0.04}$; c) $S_{0.07}$; d) $S_{0.13}$. The scale bars correspond to 200 nm.}
\end{figure}

The samples topography presented in figure~\ref{fig:AFM_topo} shows a result consistent with that of scanning electron microscopy in the main text. The wrinkles characteristic for epitaxial layers are present on all the samples. Number and size of the objects on the hBAlN layer increase with the amount of TMAl in the growth process. The shapes of those objects seem to be more and more complex for samples with higher TMAl flow. Both wrinkles and objects are reproduced figures~\ref{fig:AFM_adh},~\ref{fig:AFM_defor},~\ref{fig:AFM_stiff} that present adhesion, deformation and stiffness modulus of the material, respectively. This contrast (between flat hBAlN layer and the objects) that is observed for all AFM images proves presence of two different types of the material. To make data related to hBAlN and crystaline objects easier to compare we calculated average values of adhesion, deformation, Young's modulus. They are collected in table~\ref{tab:AFM}. The exact values of the quantities vary for each sample. However, some trends can be easily seen. The hBAlN layers adhese stronger than crystalites which can be due to different nature of bonds of surface atoms. Young's modulus values suggest that hBAlN layers have higher stiffness in comparison to objects on its surface. However, both of them deform comparably under AFM cantilever.

\begin{table}[htbp]
\centering
\caption{\label{tab:AFM}
Average values of crystalite height, adhesion, deformation and stiffness modulus of hBAlN and crystaline objects obtained with atomic force microscopy analysis for studied samples.}
%\begin{tabular}{p{5cm}|p{2cm} p{2cm} p{2cm} p{2cm}}
\begin{tabular}{l|cccc}
\hline\hline
Sample                                                                       & $S_{0.02}$ & $S_{0.04}$ & $S_{0.07}$ & $S_{0.13}$ \\
\hline \hline
crystalite height (nm)                                                       & -     & 21    & 27    & 18    \\ 
hBAlN adhesion (nN)                                                          & 1.49  & 3.89  & 4.31  & 2.98  \\
crystalite adhesion (nN)                                                     & 1.24  & 1.63  & 1.76  & 1.24  \\
hBAlN deformation (nm)                                                       & 0.66  & 1.00  & 0.60  & 0.52  \\
crystalite deformation (nm)                                                   & 1.38  & 1.10  & 0.62  & 0.42  \\
\begin{tabular}[c]{@{}l@{}}hBAlN\\ Young's modulus (GPa)\end{tabular}      & 8.8   & 7.1   & 13.0  & 14.9  \\
\begin{tabular}[c]{@{}l@{}}crystalite\\ Young's modulus (GPa)\end{tabular} & 3.9   & 6.0   & 8.0   & 10.1 \\
\hline\hline
\end{tabular}
\end{table}

\clearpage
\newpage
\section*{Raman spectroscopy of hBAlN samples}

Raman spectra were acquired using a Renishaw inVia Raman setup equipped with a 785~nm continuous wave laser excitation source and a 100$\times$ objective. Laser power density was about 300~kW/cm$^2$. For each point on the sample two spectra were collected - the first with the laser beam focused on the surface of the hBAlN layer and the second with the laser beam focused inside the substrate ($\sim$20~$\mu$m below). Then, the background spectrum was subtracted to obtain a pure epitaxial material signal.

\begin{figure}[hbtp]
\includegraphics[width=\textwidth]{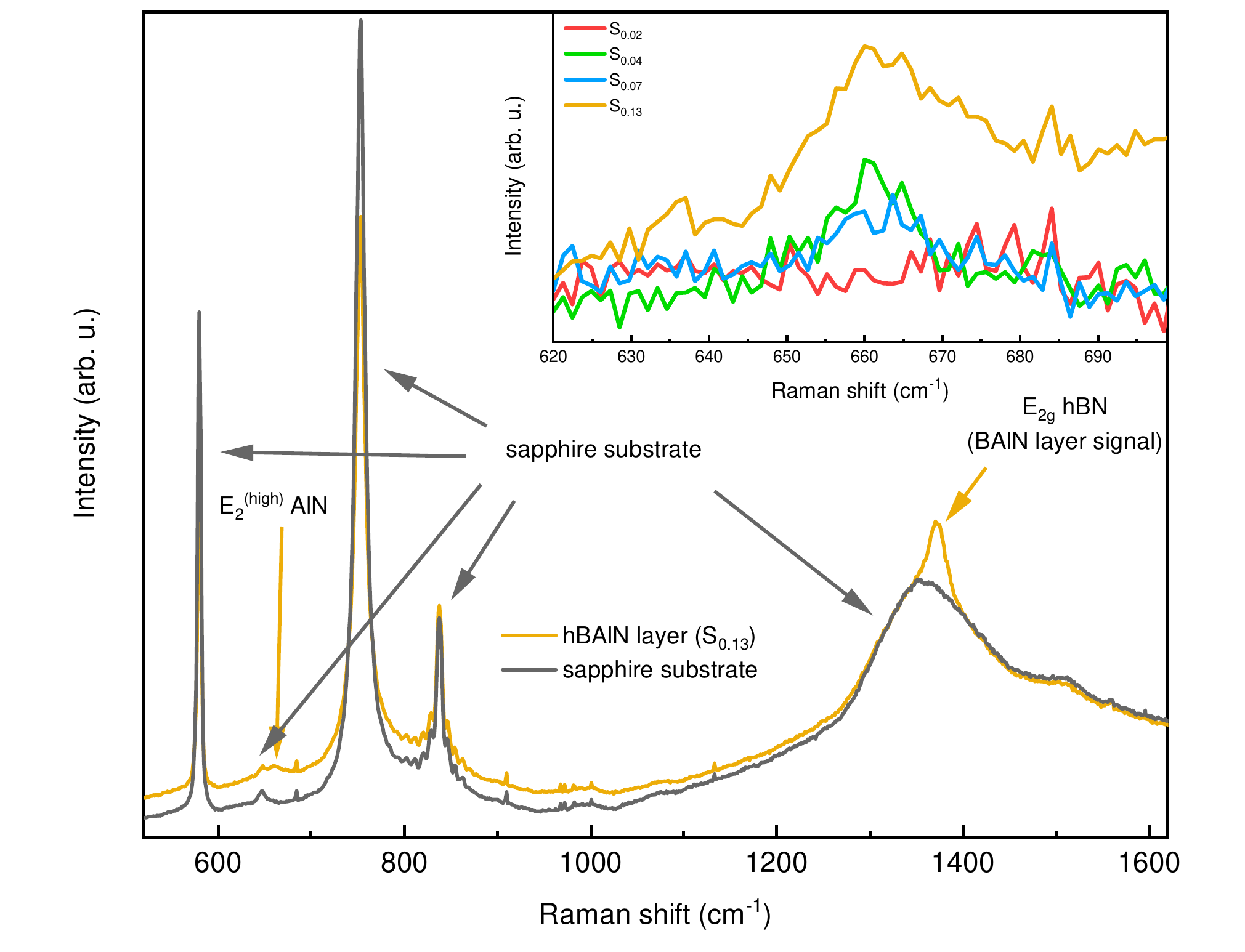}
\caption{\label{fig:Raman}Exemplary Raman spectrum collected for the sample $S_{0.13}$ with the highest TMAl:III ratio (yellow line). The spectrum collected at the same point but inside the sapphire substrate is presented by the gray line. The inset shows the zoom on the area related to $E_2^{high}$ aluminum nitride mode for the samples: $S_{0.02}$ (red line), $S_{0.04}$ (green line), $S_{0.07}$ (blue line) and $S_{0.13}$ (yellow line). The background was subtracted from the spectra presented in the inset.}
\end{figure}

In figure~\ref{fig:Raman} we present an exemplary Raman spectrum for the $S_{0.13}$ sample and the sapphire substrate spectrum collected at the same point. As can be seen, sapphire-related peaks overlap with the peaks for the hBAlN layer. By comparison of these two spectra we can distinguish two characteristic peaks coming from the hBAlN layer that are related to $E_{2g}$ phonon mode in hBN ($\sim 1366$~cm$^{-1}$ \cite{kuzuba1978}) and $E_2^{(high)}$ phonon mode in AlN ($\sim 660$~cm$^{-1}$ \cite{oliveira2001}). The mode characteristic of AlN (sp$^3$-bonded material) should not be observed for samples of hBAlN (sp$^2$-bonded material). The presence of this signal can be explained by the crystaline objects observed in SEM (figure~3 in the main text) and AFM (figure~\ref{fig:AFM_topo}-\ref{fig:AFM_stiff}) images. A closer look for this spectral region is presented in the inset of figure~\ref{fig:Raman}. When the sapphire substrate signal is substracted from the spectra of hBAlN, the AlN-related peak can be observed more clearly. The peak intensity correlates with the amount of TMAl used in the growth process. The more TMAl the peak has higher intensity. The dependence stands in analogy to the appearance of crystaline objects observed in SEM and AFM (the more TMAl, the more crystalites) as well as detected in XRD (figure~1 in the main text). All in all implies that the objects present on the surface of hBAlN layers are composed of aluminum nitride.

\clearpage
\bibliography{bibliography}